\documentstyle[12pt,amsfonts]{article}

\def\noi{\noindent}

\def\barr{\left(\begin{array}}
\def\earr{\end{array}\right)}
\def\beq#1{\begin{equation}\label{#1}}
\def\eeq{\end{equation}}
\def\ber#1{\begin{eqnarray}\label{#1} \nqq}
\def\eer{\end{eqnarray}}

\newcommand{\bear}[1]{\begin{eqnarray}\label{#1}}
\newcommand{\ear}{\end{eqnarray}}

\renewcommand{\theequation}{\arabic{section}.\arabic{equation}}
\catcode`\@=11 \@addtoreset{equation}{section}\catcode`\@=12

\newcommand{\N}{ {\mathbb N} }
\newcommand{\R}{ {\mathbb R} }

\newcommand{\sign}{\mathop{\rm sign}\nolimits}
\newcommand{\eps}{\varepsilon}
\newcommand{\p}{\partial}
\newcommand{\nn}{\nonumber}

\newcommand{\fnm}{\footnotemark}
\newcommand{\fnt}{\footnotetext}

\begin{document}

 \vspace{15pt}

 \begin{center}
 \large\bf

 On the ``scattering law'' for Kasner parameters \\
 appearing in asymptotics of an exact S-brane solution

 \vspace{15pt}

 \normalsize\bf V.D. Ivashchuk\fnm[1]\fnt[1]{rusg@phys.msu.ru}
             and  V.N. Melnikov\fnm[2]\fnt[2]{melnikov@phys.msu.su}

 \vspace{5pt}

 \it Center for Gravitation and Fundamental Metrology,
 VNIIMS, 46 Ozyornaya ul., Moscow 119361, Russia  \\

 Institute of Gravitation and Cosmology,
 Peoples' Friendship University of Russia,
 6 Miklukho-Maklaya ul.,  Moscow 117198, Russia \\

 \end{center}
 \vspace{15pt}

 \small\noi

 \begin{abstract}

    A multidimensional cosmological model with scalar and form fields
    \cite{Mel2,Mel,Mel3,Solv} is studied.
    An exact $S$-brane solution (either electric or magnetic)
    in a model with $l$ scalar fields and one antisymmetric form
    of rank $m \geq 2$ is considered. This solution is defined on
    a product manifold  containing  $n$ Ricci-flat factor spaces
    $M_1, ..., M_n$.  In the case when the kinetic term for scalar
    fields is positive definite  we singled out a special solution  governed
    by the function $cosh$. It is shown that this special solution
    has  Kasner-like asymptotics in the  limits $\tau \to  + 0$ and
    $\tau \to  + \infty$,  where $\tau$ is a synchronous time variable.
    A relation between two sets of Kasner parameters
    $\alpha_{\infty}$ and  $\alpha_{0}$ is found. This
    relation,  named as ``scattering law'' (SL) formula,
    is coinciding with the ``collision law'' (CL)
    formula obtained previously  in  \cite{Ierice} in a context of a billiard
    description of $S$-brane solutions near the singularity.
    A geometric sense of SL formula is clarified: it is shown that
    SL transformation  is a  map of a ``shadow'' part of the
    Kasner sphere $S^{N-2}$ ($N = n+l$) onto  ``illuminated'' part.
    This map  is just a
    (generalized) inversion with respect to a point $v$
    located outside the Kasner sphere $S^{N-2}$.  The shadow
    and illuminated parts of the Kasner sphere are defined
    with respect to a point-like source of light located at $v$.
    Explicit formulae for  SL transformations corresponding
    to $SM2$- and $SM5$-brane  solutions in $11$-dimensional supergravity
    are presented.

 \end{abstract}

 \vspace{20cm}

 \pagebreak

 \normalsize

\section{Introduction}

In \cite{IMb1} a multidimensional model describing the
cosmological ``evolution'' of $n$ Einstein spaces in the theory
with $l$ scalar fields and several antisymmetric forms was
considered. When electro-magnetic composite $S$-brane ansatz was
adopted, and certain restrictions on the parameters of the model
were imposed, the dynamics of the model near the singularity was
reduced to a billiard on the
 $(N-1)$-dimensional hyperbolic (Lobachevsky) space $H^{N-1}$, $N =
  n+l$.

We recall that Kasner-like solutions  have the following form
 \bear{1.m}
  g = w d\tau \otimes d\tau + \sum_{i=1}^{n} A_i
 \tau^{2 \alpha^i} g^i,
 \\   \label{1.s}
 \varphi^{\beta} =  \alpha^{\beta} \ln \tau + \varphi^{\beta}_0,
 \ear
 where
 \bear{1.k1}
  \sum_{i=1}^{n} d_i \alpha^i = 1,
  \\  \label{1.k2}
  \sum_{i=1}^{n} d_i (\alpha^i)^2 +
 \alpha^{\beta} \alpha^{\gamma} h_{\beta \gamma}= 1, \
 \ear
  $\varphi^{\beta}_0$ are constants $i = 1, \ldots, n$; $\beta,
 \gamma = 1, \ldots, l$; and $w = \pm 1$.

 It was shown in \cite{Ierice} that the set of Kasner parameters
 $(\alpha^{'A})$ after the collision with the $s$-th wall
  is defined by the Kasner set before the collision
 $(\alpha^{A})$ according to the following formula
  \beq{gcl}
   \alpha^{'A} =
               \frac{\alpha^A - 2 U^s(\alpha) U^{sA}(U^s,U^s)^{-1}}
               {1 - 2 U^s(\alpha) (U^s,U^{\Lambda})(U^s,U^s)^{-1}}.
  \eeq
  Here $(\alpha^A) =  (\alpha^i, \alpha^{\beta}) \in \R^N$, $N = n +
  l$;  $U^s$ is a brane  co-vector corresponding to the $s$-th wall and
  $U^{\Lambda}$  is a co-vector, corresponding to the $\Lambda$-term.
  These vectors and the scalar product $(.,.)$ were defined in
  \cite{IMC,IMJ,IMtop,Ierice}, see also Section 3 of this paper.

  In the special case of one
  scalar field and  $1$-dimensional factor-spaces
  (i.e.  $l= d_i =1$) this formula was suggested earlier
  in \cite{DH}. Another special case of collision law
  for multidimensional multi-scalar cosmological model
  with exponential potentials  was considered in \cite{DIMbil}.

  In this paper  we consider an exact $S$-brane solution with one brane
   (either electric or magnetic)  in the model
   with  $l$ scalar fields and one antisymmetric form
   of rank $m \geq 2$ (see Section 2) \cite{Iohta,Ifest}.
   This solution is defined on a
   product manifold  containing  $n$ Ricci-flat factor spaces.
   In Section 3 we rewrite the solution in a so-called ``minisuperspace-covariant''
   form that significantly simplifies forthcoming analysis.

   In the case when the kinetic term for scalar  fields is positive
   definite one
   we single out a special solution   governed  by the $cosh$ function.
   In Section 4 we show that this solution has a Kasner-like asymptotics in
    both limits $\tau \to  + 0$ and $\tau \to  + \infty$,
   where $\tau$ is the synchronous time variable.
   We also find a relation between two sets of Kasner parameters
   $\alpha_{\infty}$ and  $\alpha_{0}$. This relation
   (``scattering law'' formula)  is coinciding with the CL
   formula from (\ref{gcl}).

   In Section 5 we  clarify the geometric sense   of the SL.
   We have expressed the SL transformation
   in terms of a function  mapping a ``shadow'' part of the
   Kasner sphere $S^{N-2}$ onto  ``illuminated'' part. We show that this
   function  is just an inversion with respect to a point $v$
   located outside the Kasner sphere $S^{N-2}$.

    In Section 6 we present explicit formulae for SL transformations
    corresponding  to $SM2$- and $SM5$-brane
    solutions in $11$-dimensional supergravity.


 \section{$S$-brane solution }

 Here we deal with a model governed by the action
  \beq{1.1}
   S_g=\int d^Dx
   \sqrt{|g|}\biggl\{R[g]- h_{\alpha\beta}g^{MN}\p_M\varphi^\alpha
   \p_N \varphi^\beta- \frac{\theta}{m!}
   \exp[2\lambda(\varphi)]F^2\biggr\}
  \eeq
where $g=g_{MN}(x)dx^M\otimes dx^N$ is a metric,
 $\varphi=(\varphi^\alpha)\in\R^l$ is a vector of scalar fields,
 $(h_{\alpha\beta})$ is a  constant symmetric
non-degenerate $l\times l$ matrix $(l\in \N)$,
 $\theta= \pm1$,
 $F =    dA
 =  \frac{1}{m!} F_{M_1 \ldots M_{m}}
 dz^{M_1} \wedge \ldots \wedge dz^{M_{m}}$
 is a $m$-form ($m \ge1$), $\lambda$ is a 1-form on $\R^l$:
 $\lambda(\varphi)=\lambda_{\alpha} \varphi^\alpha$,
 $\alpha=1,\dots,l$. In (\ref{1.1}) we denote $|g| =   |\det
(g_{MN})|$,
 $F^2  = F_{M_1 \ldots M_{m}} F_{N_1 \ldots N_{m}} g^{M_1
         N_1} \ldots g^{M_{m} N_{m}}$.

 For pseudo-Euclidean metric of
 signature $(-,+, \ldots,+)$ one should put $\theta = 1$.

We consider $S$-brane solution (either electric or magnetic one)
to field equations corresponding to the action
 (\ref{1.1}) and depending upon one variable $u$
 \cite{Iohta,Ifest} (see also \cite{Isbr,IK,IMtop}).

 This solution is defined on the manifold
  \beq{1.2}
  M =    (u_{-}, u_{+})  \times
  M_1  \times M_2 \times  \ldots \times M_{n},
  \eeq
where $(u_{-}, u_{+})$  is  an interval belonging to $\R$, and has
the following form
 \bear{1.3}
  g= [f(u)]^{2 d(I) h/(D-2)} \biggr)
  \biggr\{ \exp(2c^0 u + 2 \bar c^0) w du \otimes du  + \\ \nn
  \sum_{i = 1}^{n} \Bigl( [f(u)]^{- 2 h  \delta^i_{I} } \Bigr)
  \exp(2c^i u+ 2 \bar c^i) g^i \biggr\}, \\ \label{1.4}
  \exp(\varphi^\alpha) =
  \left(f^{h \chi \lambda^\alpha} \right)
  \exp(c^\alpha u + \bar c^\alpha), \\ \label{1.5.9}
   F= Q  f^{- 2} du \wedge\tau(I), \quad \chi = + 1,
  \\ \label{1.5.10}
     = Q \tau(\bar I), \quad \chi = - 1,
  \ear
   $w = \pm 1$, $\alpha=1,\dots,l$.

   Here and in what follows
   \beq{1.8}
    \chi  =  +1, -1
   \eeq
 for electric or magnetic case, respectively,  $Q \neq 0$ is
 a constant (charge density parameter)
 and  $\lambda^\alpha = h^{\alpha\beta} \lambda_\beta$  where
 $(h^{\alpha\beta})=(h_{\alpha\beta})^{-1}$.

 In  (\ref{1.3})  $w = \pm 1$,
 $g^i=g_{m_i n_i}^i(y_i) dy_i^{m_i}\otimes dy_i^{n_i}$
 is a Ricci-flat  metric on $M_{i}$, $i=  1,\ldots,n$,
  \beq{1.11}
   \delta^i_{I}=  \sum_{j \in I} \delta^i_j
  \eeq
is the indicator of $i$ belonging to $I$: $\delta^i_{I}= 1$ for
 $ i \in I$ and $\delta^i_{I} = 0$ otherwise.

The set  $I = \{ i_1, \ldots, i_k \}$ is a subset of $ I_0 = \{ 1,
 \ldots,n \}$. It describes the location of $S$-brane worldvolume.
Here  and in what follows
  \beq{1.13a}
  \bar I \equiv I_0 \setminus I.
  \eeq

All manifolds $M_{i}$ are assumed to be oriented and connected and
the volume $d_i$-forms
  \beq{1.12}
  \tau_i  \equiv \sqrt{|g^i(y_i)|}
  \ dy_i^{1} \wedge \ldots \wedge dy_i^{d_i},
  \eeq
and parameters
  \beq{1.12a}
   \varepsilon(i)  \equiv {\rm sign}( \det (g^i_{m_i n_i})) = \pm 1
  \eeq
are well-defined for all $i=  1,\ldots,n$. Here $d_{i} =   {\rm
dim} M_{i}$, $i =   1, \ldots, n$;
 $D =   1 + \sum_{i =   1}^{n} d_{i}$. For any
 set $I =   \{ i_1, \ldots, i_k \} \in I_0$, $i_1 < \ldots < i_k$,
we denote
  \bear{1.13}
  \tau(I) \equiv \tau_{i_1}  \wedge \ldots \wedge \tau_{i_k},
  \\
  \label{1.15}
  d(I) \equiv   \sum_{i \in I} d_i, \\
  \label{1.15a}
  \varepsilon(I) \equiv \varepsilon(i_1) \ldots \varepsilon(i_k).
 \ear

The parameters  $h$ appearing in the solution satisfy the
relations
 \beq{1.16}
  h = K^{-1},
 \eeq
 where
 \beq{1.17}
  K =  d(I)+\frac{(d(I))^2}{2-D}+
    \lambda_{\alpha}\lambda_{\beta}
  h^{\alpha\beta}.
 \eeq

Here we assume that  $K \neq 0$.

The moduli function reads
 \bear{1.4.5}
  f(u)=
  R \sinh(\sqrt{C}(u-u_1)), \;
  C > 0, \; K \eps <0; \\ \label{1.4.7}
  R \sin(\sqrt{|C|}(u-u_1)), \;
  C<0, \; K \eps < 0; \\ \label{1.4.8}
  R \cosh(\sqrt{C}(u-u_1)), \;
  C > 0, \; K \eps > 0; \\ \label{1.4.9}
  |Q||K|^{1/2}(u-u_1), \; C=0, \; K \eps<0,
  \ear
 where $R = |Q|| K/ C|^{1/2}$, and $C$, $u_1$  are constants.

 Here   $\eps = \eps(I) \theta $ for electric case
 and $\eps = -\eps[g] \eps(I) \theta$  for magnetic case,
 where $\eps[g] = \sign\det(g_{MN})$.

 Vectors $c=(c^A)= (c^i, c^\alpha)$ and
 $\bar c=(\bar c^A)$ obey the following constraints
 \beq{1.27}
  \sum_{i \in I} d_i c^i - \chi \lambda_{\alpha} c^\alpha=0,
  \qquad
  \sum_{i \in I}d_i \bar c^i- \chi \lambda_{\alpha} \bar c^\alpha = 0,
   \eeq
  \bear{1.30aa}
   c^0 = \sum_{j=1}^n d_j c^j,
  \qquad
   \bar  c^0 = \sum_{j=1}^n d_j \bar c^j,
  \\  \label{1.30a}
   C  K^{-1} +
    h_{\alpha\beta}c^\alpha c^\beta+ \sum_{i=1}^n d_i(c^i)^2
  - (\sum_{i=1}^nd_ic^i)^2 = 0.
 \ear

Due to (\ref{1.5.9}) and  (\ref{1.5.10}), the dimension of brane
worldvolume $d(I)$ is defined by
 \beq{1.16a}
  d(I)=  m -1, \quad d(I)=   D- m -1,
 \eeq
for electric and magnetic cases, respectively. For $Sp$-brane we
have $p = p(I) = d(I)-1$. The solution under consideration is a
special one  brane case of intersecting $S$-brane solutions from
 \cite{Isbr,Iohta,Ifest}.

 \section{Minisuperspace-covariant notations}

 Our solution may be written also in the
 so-called ``minisuperspace-covariant'' form following from
 the sigma-model solution \cite{IK}.

 The metric (\ref{1.3}) has the structure
 \beq{2.m}
   g= e^{2\gamma_0(u)} du \otimes du + \sum_{i=1}^{n} e^{2\phi^i(u)} g^i,
 \eeq
 where
 \beq{2.g}
  \gamma_0 = \sum_{i=1}^{n} d_i \phi^i(u).
 \eeq
 Introducing a
 collective variable  $x = (x^{A})=(\phi^i,\varphi^\alpha)$ we get
 a minisuperspace-covariant relation (see (\ref{1.3})
 and(\ref{1.4})):
  \beq{2.0}
   x^A(u)= - \frac{U^{A}}{(U,U)}\ln |f(u)| + c^A u +  \bar{c}^A,
  \eeq
  where the function $f(u)$ was defined in
  (\ref{1.4.5})-(\ref{1.4.9}) and $c=(c^A)= (c^i, c^\alpha)$.

 The linear and quadratic constraints from (\ref{1.27})
 and (\ref{1.30a}), respectively, read in a
 minisuperspace covariant form as:
 \beq{2.8}
   U_A c^A = 0, \qquad U_A \bar{c}^A = 0,
 \eeq
 and
  \beq{2.10}
   \frac{C}{(U,U)}  + \bar G_{AB} c^A c^B = 0.
  \eeq

 Here
 \beq{2.3}
  (U_A) =  (d_i \delta^i_{I}, -\chi \lambda_{\alpha}),
  \eeq
 is the so-called brane co-vector (U-vector)  \cite{IMC,IMJ}
  \beq{2.3a}
  (U^{A})= (\bar{G}^{AB} U_B) = (\delta^i_{I} - \frac{d(I)}{D-2},
  - \chi \lambda^\alpha),
  \eeq
  and
  \beq{2.1}
  (\bar{G}_{AB})=\barr{cc}
  G_{ij}& 0\\
  0& h_{\alpha\beta}
  \earr,
  \qquad
 (\bar G^{AB})=\left(\begin{array}{cc}
  G^{ij}&0\\
 0&h^{\alpha\beta}
 \end{array}\right)
  \eeq
 are, correspondingly,  a minisuperspace metric
 and inverse to it, where (see \cite{IMZ})
 \beq{2.2}
   G_{ij}= d_i \delta_{ij} - d_i d_j, \qquad
   G^{ij}=\frac{\delta^{ij}}{d_i}+\frac1{2-D}.
 \eeq

In what follows we use a scalar product \cite{IMC}
 \beq{2.4}
  (U,U')=\bar G^{AB} U_A U'_B,
 \eeq
for $U = (U_A), U' = (U'_A) \in \R^N$, $N = n + l$.

 In (\ref{2.10}) we used the relation
 \beq{2.7}
  (U,U)= K.
 \eeq

 The logarithm of the lapse function (\ref{2.g})
 may be also  written in the minisuperspace covariant form
 \beq{2.gd}
  \gamma_0 = U^{\Lambda}_A x^A,
 \eeq
 where
  \beq{2.l}
   (U^{\Lambda}_A) = (d_i, 0)
  \eeq
  is $U$-vector, corresponding to the $\Lambda$-term
  \cite{IMC,IMJ}.

  We will use also the relations
   \beq{2.c}
    c^0 = U^{\Lambda}_A c^A,
   \eeq
  and
  \bear{2.ul}
   (U,U^{\Lambda}) = - \frac{d(I)}{D-2},
   \\ \label{2.ll}
   (U^{\Lambda},U^{\Lambda}) = - \frac{D-1}{D-2}.
  \ear

 \section{Scattering law for Kasner parameters}

  Here we restrict our consideration by a special solution
  with   $K  = (U,U) > 0$, $C >0$ and $\eps > 0$. We also
  put  the matrix $(h_{\alpha\beta})$ to be positive definite.

  In this case the solution  is governed  by the
  moduli function  $f(u)= R \cosh(\sqrt{C}(u-u_1))$
  and is defined for all $u \in (- \infty, + \infty)$.

\subsection{Kasner-like behaviour}

 Let us consider our solution in a synchronous time:
 \beq{3.tau}
  \tau = \eta  \int ^{u}_{u_0} d \bar{u} e^{\gamma_0( \bar{u})},
 \eeq
 where $\eta = \pm 1$,  $u_{0}$ is constant and
  \beq{3.l}
   e^{\gamma_0(u)} = |f(u)|^{d(I) h/(D-2)} \exp(c^0 u  + \bar c^0)
  \eeq
 is a lapse function.

 Due to
 \beq{3.fas}
   f \sim \frac{R}{2} \exp( \pm \sqrt{C}(u - u_1)),
 \eeq
  for $u \to \pm \infty$,
  we get asymptotical relations for the lapse function
  \beq{3.flas}
   e^{\gamma_0} \sim {\rm const} \exp( b_{\pm} \sqrt{C} u ),
  \eeq
  as $u \to \pm \infty$,  with
  \beq{3.1}
    b_{\pm} = \pm \frac{h d(I)}{D-2} +  \frac{c^0}{\sqrt{C}}.
  \eeq

  Using relations (\ref{2.c}), (\ref{2.ul})  and $h = (U,U)^{-1}$
  we could rewrite the parameters  $b_{\pm}$ in a minisuperspace-covariant form
   \beq{3.b}
     b_{\pm} = \mp \frac{(U^{\Lambda},U)}{(U,U)} +  (s,U^{\Lambda}).
   \eeq
  where
  \beq{3.s}
    s  = (s_A) = (\bar G_{AB} c^B/\sqrt{C}),
  \eeq
  is a co-vector, obeying relations:
   \bear{3.2}
       (s,U) = 0,
   \\ \label{3.3}
     \frac{1}{(U,U)}  + (s,s) = 0.
  \ear
  following just from  (\ref{2.8}) and (\ref{2.10}).  In
  derivation of (\ref{3.b}) we used the relation
   \beq{3.c}
    c^0  = (s,U^{\Lambda}) \sqrt{C},
   \eeq
   following from (\ref{2.c}) and (\ref{3.s}).

   In what follows we will use an inequality
   \beq{3.4}
    |(s,U^{\Lambda})| >  \frac{|(U^{\Lambda},U)|}{(U,U)} > 0,
   \eeq
    proved in Appendix. The proof uses relations
   (\ref{3.2}), (\ref{3.3}) and $(U,U) > 0$.

   The parameter $c^0$ is a non-zero one (otherwise the relation (\ref{1.30a})
   would be incompatible with the conditions $C >0$, $K > 0$ and
   positive definiteness of the matrix $(h_{\alpha\beta}))$.
   It follows from inequality (\ref{3.4}) that $b_{\pm}$
   are also non-zero and
    \beq{3.5}
    \sign(b_{\pm}) = \sign((s,U^{\Lambda})) = \sign(c^0).
   \eeq

   It may be verified that due to (\ref{3.4}) the lapse function $e^{\gamma_0(u)}$
   is monotonically increasing from $+0$ to $+ \infty$ for $c^0 >0$
   and monotonically decreasing from $+ \infty$ to $+0$ for $c^0 <
   0$.

     We define synchronous time variable to be
     \beq{3.t1}
      \tau =   \int^{u}_{- \infty} d \bar{u} e^{\gamma_0( \bar{u})},
     \eeq
   for $c^0 >0$ and
     \beq{3.t2}
      \tau =   \int^{+ \infty}_{u} d \bar{u} e^{\gamma_0( \bar{u})},
     \eeq
   for $c^0 < 0$. Then, the function $\tau = \tau(u)$
   is monotonically increasing from $+0$ to $+ \infty$ for $c^0 >0$
   and monotonically decreasing from $+ \infty$ to $+0$ for $c^0 <
   0$.

     We have the following asymptotical relations for $\tau = \tau(u)$
    \beq{3.6}
    \tau \sim {\rm const} \  b_{\pm}^{-1} \exp( b_{\pm} \sqrt{C} u ),
    \eeq
    as $u \to \pm \infty$.

    For the collective variable
    $(x^{A})=(\phi^i,\varphi^\alpha)$ from (\ref{2.0})
    we get (see (\ref{3.fas}))
   \beq{3.7}
    x^A(u) \sim  \mp \frac{U^{A} \sqrt{C} u}{(U,U)} + c^A u +  \hat{c}^A,
   \eeq
   as $u \to \pm \infty$, where $\hat{c}^A$ are constants.
   Hence, due to (\ref{3.6}) we are led to Kasner-like
   asymptotics  written in a minisuperspace covariant form
    \beq{3.8}
     x^A \sim  \alpha^A_{\pm} \ln \tau +  x^A_{\pm},
    \eeq
     for $u \to \pm \infty$, where $x^A_{\pm}$ are constants and
    \beq{3.9}
       \alpha^A_{\pm} =  [\mp \frac{U^{A}}{(U,U)} +  s^{A}]/b_{\pm}
    \eeq
    are collective Kasner-like parameters,  corresponding to $u \to \pm \infty$.

    Asymptotical relations (\ref{3.8}) could be also rewritten
    in the form assigned to proper time asymptotics, i.e.
    \bear{3.10a}
     x^A \sim  \alpha^A_{0} \ln \tau +  x^A_0, \ {\rm as} \ \tau \to
     +0,    \\ \label{3.10b}
     x^A \sim  \alpha^A_{\infty} \ln \tau +  x^A_{\infty}, \ {\rm as} \ \tau \to
     + \infty.
    \ear
     Here
     \beq{3.11a}
      \alpha^A_{0} = \alpha^A_{-}, \quad \alpha^A_{\infty} = \alpha^A_{+}
     \eeq
   for $c^0 >0$ and
      \beq{3.11b}
      \alpha^A_{0} = \alpha^A_{+}, \quad \alpha^A_{\infty} = \alpha^A_{-}
     \eeq
     for $c^0 < 0$ and $x^A_{0}$, $x^A_{\infty}$ are constants.

  It follows from definitions of Kasner parameters (\ref{3.9}) that
   \bear{3.12}
    \bar G_{AB}\alpha^A_{\pm} \alpha^B_{\pm}   = 0, \\
    \label{3.13}
    U(\alpha_{\pm}) = U_A \alpha^A_{\pm}  = \mp \frac{1}{b_{\pm}}, \\
    \label{3.14}
    U^{\Lambda}(\alpha_{\pm})  =  1,
    \ear
    see (\ref{3.b}), (\ref{3.2}) and (\ref{3.3}).

   In components relations (\ref{3.12}) and  (\ref{3.14})
   read as
   \beq{3.15k1}
    \sum_{i=1}^{n} d_i \alpha^i_{\pm} =
    \sum_{i=1}^{n} d_i (\alpha^i_{\pm})^2 +
   \alpha^{\beta}_{\pm} \alpha^{\gamma}_{\pm} h_{\beta \gamma}= 1. \
    \eeq

    Thus, we are led to Kasner-like relations (\ref{1.k1}) and
    (\ref{1.k2})
    for $\alpha_{\pm} = (\alpha^A_{\pm})$.
    Hence, $\alpha_{0} = (\alpha^A_{0})$ and $\alpha_{\infty} = (\alpha^A_{\infty})$
    also obey relations (\ref{1.k1}) and (\ref{1.k2}).

    So, we
    obtain a Kasner-like asymptotical behaviour of our special solution
    (with $C >0$ and $\eps > 0$)  for i) $\tau \to +0$ and for
    ii) $\tau \to   + \infty$, as well.  The Kasner-like  behaviour in the case i)
    is in agreement with the general result from \cite{IMb1}.

    Using  (\ref{3.5}) and (\ref{3.13}) we get
       \bear{3.16i}
       U(\alpha_{0}) = U_A  \alpha^A_{0}  = \sum_{i \in I} d_i \alpha_{0}^i
                         - \chi \lambda_{\beta} \alpha_{0}^\beta > 0,
       \\ \label{3.16ii}
       U(\alpha_{ \infty}) = U_A \alpha^A_{\infty}  =
       \sum_{i \in I} d_i \alpha_{\infty}^i
               - \chi \lambda_{\beta} \alpha_{\infty}^\beta  < 0.
       \ear
       The first inequality  (\ref{3.16i}) is a special case of a set of inequalities
       derived in  \cite{IMb1}. Any inequality of such type  corresponds to a billiard
       wall in hyperbolic (e.g. Lobachevsky) space.

   \subsection{Scattering law}

          Now we derive a relation between Kasner sets  $\alpha_{0}$
  and  $\alpha_{\infty}$.

   We start with the formulas:
    \beq{3.17}
     b_{+} \alpha_{+} - b_{-} \alpha_{-} =  -  \frac{2 U}{(U,U)}
    \eeq
   and
   \beq{3.18}
     b_{+}  - b_{-} =  -  \frac{2(U^{\Lambda},U)}{(U,U)}
    \eeq
   following from (\ref{3.9}) and  (\ref{3.b}), respectively.
    Using these relations  and (\ref{3.13})  we get
    \beq{3.19}
      \alpha_{\pm}^A =
               \frac{\alpha_{\mp}^A - 2 U^A U(\alpha_{\mp}) (U,U)^{-1}}
               {1 - 2 U(\alpha_{\mp}) (U,U^{\Lambda})(U,U)^{-1}} .
     \eeq
     This formula gives a scattering law formula for Kasner parameters
     (see definitions (\ref{3.11a}) and (\ref{3.11b}))

    \beq{3.20}
      \alpha_{ \infty} =
               \frac{\alpha_{ 0} - 2 U U(\alpha_{ 0}) (U,U)^{-1}}
               {1 - 2 U(\alpha_{ 0}) (U,U^{\Lambda})(U,U)^{-1}} = S(\alpha_0).
     \eeq
    coinciding for $U = U^s$ with the collision law formula (\ref{gcl})
     derived in \cite{Ierice}.

    It should be also noted that due to (\ref{3.19})  the inverse function $S^{-1}$
    is given by just the same relation
     \beq{3.20a}
      \alpha_{0} =
               \frac{\alpha_{\infty} - 2 U U(\alpha_{\infty}) (U,U)^{-1}}
               {1 - 2 U(\alpha_{\infty }) (U,U^{\Lambda})(U,U)^{-1}} = S^{-1}(\alpha_{\infty}).
     \eeq

 \section{Geometric meaning of the scattering law}

 Let us clarify the geometric meaning of the scattering law.
 Since the matrix $(h_{\alpha \beta})$ is positive definite
 the Kasner relations (\ref{1.k1}) and (\ref{1.k2}) describe
 an ellipsoid isomorphic to a unit $(N-2)$-dimensional sphere
 $S^{N-2}$ which is a subset of  $\R^{N-1}$, $N=n+l$.
 Thus,  the sets of Kasner parameters $\alpha$
 may be parametrized  by vectors $\vec{n}  \in S^{N-2}$, i.e.
  $\alpha= \alpha(\vec{n})$.
   Let us show that the scattering law formula (\ref{gcl}) in terms
   of $\vec{n}$-vectors reads as follows
   \beq{4.gcl}
   \vec{n'} = \frac{(\vec{v}^2 - 1)\vec{n}
   + 2(1 - \vec{v}\vec{n})\vec{v}}{(\vec{v} - \vec{n})^2}
   \eeq
   where $\vec{v}$ is a vector belonging to $\R^{N-1}$ with
   $|\vec{v}| > 1$.

 Now we proceed with derivation of (\ref{4.gcl}).
 The minisuperspace metric (\ref{2.1}) has a pseudo-Euclidean
 signature $(-,+, \ldots ,+)$, since the matrix $(G_{ij})$ has the
 pseudo-Euclidean signature \cite{IMZ} and $(h_{\alpha \beta})$ has the
 Euclidean one. There exists a linear transformation
 \beq{4.1}
  z^{a}=e^{a}_{A}x^{A},
 \eeq
 diagonalizing the minisuperspace metric (\ref{2.1})
  \beq{4.2}
  \bar{G}_{AB} = \eta_{ab} e^{a}_{A} e^{a}_{B}
  \eeq
 where here and in what follows
 \beq{4.3}
 (\eta_{ab})=(\eta^{ab}) = diag(-1,+1, \ldots,+1),
 \eeq
 $(e^{a}_{A})$ is a non-degenerate matrix of  linear transformations
 and $a,b = 0, \ldots,N-1$.

 The matrix  $(e^{a}_{A})$  satisfies the relation
 \beq{4.5}
  \eta^{ab} =
   e^{a}_{A}\bar{G}^{AB} e^{b}_{B} = (e^{a},e^{b}),
  \eeq
  equivalent to (\ref{4.2}) where here $e^a = (e^a_A)$.

 For the inverse matrix
 $(e_{a}^{A}) = (e^{a}_{A})^{-1}$ we obtain from (\ref{4.5})
  \beq{4.9}
   e_{a}^{A}    = \bar{G}^{AB} e^{b}_{B} \eta_{ba}.
  \eeq

 Here we put as in \cite{IMb1}
  \beq{4.10}
  e^{0} = q^{-1} U^{\Lambda},
  \eeq
 where
  \beq{4.10q}
  q =  [- (U^\Lambda,U^\Lambda)]^{1/2} = [(D-1)/(D-2)]^{1/2}.
  \eeq
  We remind that $U^{\Lambda}$  is
  a time-like co-vector, i.e. $(U^{\Lambda},U^{\Lambda}) < 0$.
  (Explicit relations for other  co-vectors $e^i$,
  $i = 1, \ldots, N-1$, are irrelevant for our consideration.
  A possible  choice of  $e^i$ is the following one: for $i = 1, \ldots, n-1$,
   these co-vectors could be found from relations for $z^i$ of Ref. \cite{IMZ},
   while for $i = n, \ldots, N-1$ these co-vectors could be readily obtained by
   diagonalization of the matrix $(h_{\alpha \beta})$.)

  Let us define a ``frame'' U-vector:
 \beq{4.11}
 \hat{U}_a  = e_{a}^{A} U_A = (U, e^{b}) \eta_{ba},
 \eeq
 (see (\ref{4.9})).
  It follows from (\ref{2.ul}), (\ref{4.10}) and (\ref{4.11})
  that
 \beq{4.12}
  \hat{U}_0  = - (U, e^{0}) = - q^{-1} (U,U^{\Lambda}) > 0.
 \eeq

  We also define ``frame'' $\alpha$-parameters:
  \beq{4.15}
  \hat{\alpha}^a  = e^a_{A} \alpha^A.
  \eeq
  In terms of $\hat{\alpha}$-parameters the
  Kasner relations (written in a ``minisuperspace covariant'' form)
   \bear{4.16}
    \bar G_{AB}\alpha^A \alpha^B  = 0,
    \\     \label{4.17}
    U^{\Lambda}(\alpha)  = 1
    \ear
  read as
    \bear{4.16f}
    \eta_{ab} \hat{\alpha}^a \hat{\alpha}^b  =
    - (\hat{\alpha}^0)^2  + \sum_{i =1}^{N-1} (\hat{\alpha}^i) = 0 ,
    \\     \label{4.17f}
    U^{\Lambda}(\alpha) = q e^{0}(\alpha) = q e^0_{A} e^{A}_b
    \hat{\alpha}^b = q \hat{\alpha}^0 = 1.
    \ear
   These equations imply \cite{IMb1}
   \beq{4.18}
   \hat{\alpha}^0 = q^{-1}, \qquad \hat{\alpha}^i = q^{-1} n^i,
   \eeq
   $i = 1, \ldots, N-1$, where the vector $\vec{n} = (n^i) \in \R^{N-1}$
   has the unit length: $\vec{n}^2 = 1$, i.e. $\vec{n}  \in S^{N-2}$.
   Relations (\ref{4.15}) and  (\ref{4.18}) define a one-to-one correspondence
   between the points of the Kasner sphere $S^{N-2}$ and  Kasner
   sets $\alpha$.

    Now  we define a  vector    $\vec{v} = (v_i) \in \R^{N-1}$
    by formula
    \beq{4.19}
      v_i  = - \hat{U}_i/\hat{U}_0,
    \eeq
    $i = 1, \ldots, N-1$ \cite{IMb1}. Since
    \beq{4.20}
   \eta^{ab} \hat{U}_a \hat{U}_b =
     - (\hat{U}_0)^2 +  \sum_{i =1}^{N-1} (\hat{U}_i)^2 = (U,U) >
     0,
    \eeq
    we get $|\vec{v}| > 1$. Using relations (\ref{4.11}),
   (\ref{4.15}), (\ref{4.18})  and  (\ref{4.19}),
    we obtain
     \beq{4.21}
      U(\alpha) = U_A \alpha^A = \hat{U}_a \hat{\alpha}^a =
      q^{-1} \hat{U}_0 (1 - \vec{v}\vec{n}).
    \eeq

    Due to (\ref{4.21}) and $\hat{U}_0 > 0$, the
    following equivalences take place
    \bear{4.22a}
    U(\alpha) > 0 \Leftrightarrow  \vec{v}\vec{n} < 1,
    \\  \label{4.22b}
    U(\alpha) < 0 \Leftrightarrow  \vec{v}\vec{n} > 1.
    \ear
    Since the asymptotical Kasner sets
    $\alpha_{\infty} = \alpha(\vec{n}_{\infty})$
    and $\alpha_{0} = \alpha(\vec{n}_{0})$ obey the inequalities
    $U(\alpha_{\infty}) < 0$ and $U(\alpha_{0}) > 0$, we
    get
     \beq{4.23}
      \vec{v}\vec{n}_{0} < 1, \qquad \vec{v}\vec{n}_{\infty} > 1.
     \eeq
    Geometrically, this means that the endpoint of the vector
    $\vec{n}_{0}$ is not illuminated by a point-like source
    of light located at the endpoint of the vector $\vec{v}$
    (i.e. the endpoint of $\vec{n}_{0}$ belongs to the ``shadow side''
     of the Kasner sphere),
    while the endpoint of the vector $\vec{n}_{\infty}$ is
    illuminated by this source.

    Using the definitions (\ref{4.11}) and (\ref{4.15})
    we rewrite the scattering law formula (\ref{gcl})
    (with $U^s = U$) in a (equivalent) ``frame representation''
     \beq{4.24}
     \hat{\alpha}^{'a} =
        \frac{\hat{\alpha}^a - 2 \hat{U}(\hat{\alpha}) \hat{U}^{a}(U,U)^{-1}}
        {1 - 2 \hat{U}(\hat{\alpha}) (U,U^{\Lambda})(U,U)^{-1}}.
     \eeq
    where $\hat{U}(\hat{\alpha}) = \hat{U}_a \hat{\alpha}^a = U(\alpha)$
    is given by (\ref{4.21}) and
      \beq{4.25}
      \hat{U}^{a} =  e^a_{A} U^A = \eta^{ab} \hat{U}_{b}.
      \eeq

     Formula (\ref{4.24}) is  satisfied identically for $a =0$,
     due to relations (\ref{4.12}), (\ref{4.18}) and $\hat{U}^{0} = -
     \hat{U}_{0}$. It may be verified using
    (\ref{4.12}), (\ref{4.18})-(\ref{4.21}) and $\hat{U}^{i} =
     \hat{U}_{i}$, $i > 0$, that  for $a = i >0$ relation (\ref{4.24})
     coincides with (\ref{4.gcl}). Thus, we proved  the
     formula (\ref{4.gcl}) with  $\vec{n} = \vec{n}_{0}$
     and $\vec{n'} =  \vec{n}_{\infty}$.

     It follows from (\ref{4.gcl}) that
     \beq{4.26}
     \vec{n'} - \vec{v} = B (\vec{n} - \vec{v}),
     \eeq
     where $B = (\vec{v}^2 - 1)/(\vec{v} - \vec{n})^2 >0$.
     Thus, the endpoints of the vectors $\vec{v}$, $\vec{n'}$
     and $\vec{n}$ belong to one line. Hence, the endpoint
     of the vector $\vec{n'}$ may be obtained  as a point of
     intersection   of the Kasner sphere with the line connecting
     the endpoints of the vectors $\vec{v}$ and $\vec{n}$.
     Thus, the scattering law transformation (\ref{4.gcl})
     is just an inversion with respect to a point $v$ located
         outside the Kasner sphere $S^{N-2}$. The point
     $v$ is the endpoint of the vector $\vec{v}$. This
     transformation map the ``shadow domain'' of Kasner
     sphere onto ``illuminated domain''.

     {\bf Remark.} We remind that according to
     an analysis carried out in \cite{IMb1} the solution
     under consideration is described  for $\tau \to 0$
     by a geodesic motion of a point-like particle in a
     billiard. This billiard belongs to the
       Lobachevsky space $H^{N-1}$ identified
     with the unit ball $D^{N-1} = \{\vec{y}: |\vec{y}| < 1  \}$.
     It is described by the  relation: $|\vec{y} - \vec{v}| > r$,
     where $r = \sqrt{\vec{v}^2 - 1}$. The billiard wall is
     a part (belonging to $D^{N-1}$) of a $(N-2)$-dimensional sphere with
     a center in the endpoint of $\vec{v}$ and  radius  $r$.

    \section{Example: $D =11$ supergravity}

   Now we consider, as an example, $D =11$
   supergravity \cite{CJS}. The bosonic sector action reads
   in this case as
    \beq{5.8a}
    S = \int d^{11}z \sqrt{|g|} \biggl\{R[g] - \frac{1}{4!} F^2 \biggr\}
      +  c \int_{M} A \wedge F \wedge F
    \eeq
 where  $F = d A$ is a 4-form and constant $c$ is irrelevant for our
 consideration. Since the second term in (\ref{5.8a})
 (called as Chern-Simons term) does not
 depend upon a metric, the Hilbert-Einstein equations are not changed when
 it is omitted . The only modification  of  equations  of motion is
 related to Maxwell-type equations
  \beq{6.43}
   d*F = {\rm const} \ F \wedge F.
  \eeq
 Due to relations for $F$ in (\ref{1.5.9}) and  (\ref{1.5.10})
 we get $F \wedge F = 0$. Thus, the solution from the Section 2
 in the special case $D=11$, $w = -1$, $l =0$ (i.e. when scalar fields are
 absent) and  $m =4$ gives us either $SM2$- or $SM5$-brane  solution in
  $11$-dimensional supergravity (see also \cite{S2} and references therein).

   In this case $\alpha = (\alpha^i)$
    and the relations on Kasner parameters (\ref{1.k1}) and (\ref{1.k2})
    read
   \beq{3.k}
   \sum_{i=1}^{n} d_i \alpha^i =  \sum_{i=1}^{n} d_i (\alpha^i)^2  = 1.
   \eeq

    For $U  = (U_i)$ we get $(U,U) = 2$, $(U,U^{\Lambda}) = -
    \frac{1}{9} d(I)$  and $U^i = \delta^i_{I} - \frac{1}{9} d(I)$.
    Here $d(I) = 3$ for electric $SM2$-brane and
        $d(I) = 6$ for magnetic $SM5$-brane.

   Thus,  the scattering law (\ref{3.20}) for $SM$-branes reads as
   follows
     \beq{3.sga}
      \alpha_{ \infty}^i =
               \frac{\alpha_{0}^i - (\delta^i_{I} - \frac{1}{9} d(I)) U(\alpha_{0})}
               {1 +   \frac{1}{9} U(\alpha_{ 0})  d(I)} .
     \eeq
     where
     \beq{3.sgb}
      U(\alpha_{0}) = \sum_{i \in I} d_i \alpha_{0}^i > 0.
     \eeq

    In terms of Kasner sphere parametrization of
    $\alpha$-parameters   the scattering law
    relation (\ref{3.sga}) is given by the formula (\ref{4.gcl})
    with $\vec{n} = \vec{n}_{0}$, $\vec{n'} = \vec{n}_{\infty}$
    and $N =n$.

      Using $(U,U) = 2$ and (\ref{4.12}) we get
      $\vec{v}^2 = 21$ in the electric case  and
      $\vec{v}^2 = 6$ in the magnetic case \cite{IMb1}. This means that
      illuminated part of the Kasner sphere  $S^n$ (containing
      endpoints of $\vec{n}_{\infty}$) is larger in the
      electric case, while the shadow domain (containing
      endpoints of $\vec{n}_{0}$) is larger in the magnetic case.

   \section{Conclusions}

   We have considered the exact $S$-brane solution with one brane
   (either electric or magnetic) to field equations corresponding to
    the action (\ref{1.1}) containing  $l$ scalar fields and one antisymmetric form
   of rank $m \geq 2$ \cite{Iohta,Ifest}. This solution is defined on the
   product manifold (\ref{1.2}) containing  $n$ Ricci-flat factor spaces
   $M_1, ..., M_n$.

   In the case when the matrix $(h_{\alpha\beta})$ is positive definite
   we have singled out a special solution   governed  by  $cosh$ moduli
   function. We have shown that this solution has Kasner-like asymptotics in the
   limits $u \to \pm \infty$, where $u$ is the harmonic time variable,
   or, equivalently, in the limits $\tau \to  + 0$ and $\tau \to  + \infty$,
   where $\tau$ is the synchronous  time variable.

   We have found a relation between two sets of Kasner parameters
   $\alpha_{\infty}$ and  $\alpha_{ 0}$. Remarkably, the
   relation between them  $\alpha_{\infty} = S(\alpha_{ 0})$
   is coinciding with the ``collision law'' formula from
   \cite{Ierice}.
    We have also clarified the geometrical sense   of the scattering law.
    Namely, we have expressed the scattering law transformation
    in terms of a function  mapping a ``shadow'' part of the
    Kasner sphere $S^{N-2}$ onto  ``illuminated'' one. This
    function  is just an inversion with respect to a point $v$
    located outside the Kasner sphere $S^{N-2}$. The shadow
    and illuminated parts of the Kasner sphere are defined
    w.r.t.  a point-like source of light located at $v$.

    We have also written explicit formulae for
    scattering law transformations corresponding
    to $SM2$- and $SM5$-brane solutions
    in $11$-dimensional supergravity.

 \renewcommand{\theequation}{\Alph{section}.\arabic{equation}}
 \renewcommand{\thesection}{}
 \setcounter{section}{0}

 \newpage

 \section{Appendix}

  Let us prove the inequality (\ref{3.4})

    $$|(s,U^{\Lambda})| >  \frac{|(U^{\Lambda},U)|}{(U,U)} > 0,$$

  for a vector $s = (s^A) \in \R^N$ ($N = n + l$) obeying relations
    $(s,U) = 0$,  $(s,s) =  - 1/(U,U)$.
  Here the scalar-product    $(U,U')=\bar G^{AB} U_A U'_B$,
  is defined by the matrix  $(\bar G^{AB})$ from (\ref{2.1})
  with a positive definite matrix $(h_{\alpha\beta})$ and
  $G^{ij}= \delta^{ij}d_i^{-1} + (2-D)^{-1}$. We also use here
  the following relations  $(U,U) > 0$, $(U^{\Lambda}_A) = (d_i, 0)$ and
   $(U^{\Lambda},U^{\Lambda}) < 0$.

   {\bf Proof.} Let us define the vector
    \beq{A.1}
    U_1 = U -  \frac{(U,U^{\Lambda})}{(U^{\Lambda},U^{\Lambda})}
    U^{\Lambda}.
    \eeq
    It is clear that $(U_1,U^{\Lambda})  = 0$ and
    \beq{A.2}
    (U_1,U_1) = (U,U) -
    \frac{(U,U^{\Lambda})^2}{(U^{\Lambda},U^{\Lambda})}> 0.
    \eeq
     since  $(U,U) > 0$ and  $(U^{\Lambda},U^{\Lambda}) < 0$.
     Let us define vectors:
     \bear{A.3}
     s_0 =  \frac{(s,U^{\Lambda})}{(U^{\Lambda},U^{\Lambda})}
     U^{\Lambda},
    \\ \label{A.4}
     s_1 =  \frac{(s,U_1)}{(U_1,U_1)} U_1,
     \\ \label{A.5}
      s_2 = s - s_0 - s_1.
      \ear
     $s_0$, $s_1$ and $s_2$ are mutually orthogonal
    and hence
    \beq{A.6}
       (s,s) = (s_0,s_0) + (s_1,s_1) + (s_2,s_2).
      \eeq

     For the  first two terms in r.h.s. of (\ref{A.6})
     we get
     \bear{A.7}
       (s_0,s_0) =
       \frac{(s,U^{\Lambda})^2}{(U^{\Lambda},U^{\Lambda})},
       \\ \label{A.8}
       (s_1,s_1) = \frac{(s,U_1)^2}{(U_1,U_1)} =
        \frac{(s,U^{\Lambda})^2}{(U^{\Lambda},U^{\Lambda})}
        \frac{(U,U^{\Lambda})^2}{[(U,U)(U^{\Lambda},U^{\Lambda}) -(U,U^{\Lambda})^2]}
      \ear
       that implies
      \beq{A.9}
       (s,s) =
        \frac{(s,U^{\Lambda})^2
        (U,U)}{(U,U)(U^{\Lambda},U^{\Lambda})
        -(U,U^{\Lambda})^2} + (s_2,s_2).
      \eeq
      For the third term in r.h.s. of (\ref{A.6}) the following
      inequality is valid
       \beq{A.10}
       (s_2,s_2) \geq 0,
      \eeq
       Indeed, due to $(s_2,U^{\Lambda})= 0$, or, equivalently,
       $\sum_{i = 1}^n s_2^i d_i = 0$ , and the
       positive definiteness of the  matrix
       $(h_{\alpha\beta})$, we obtain
       \beq{A.11}
        (s_2,s_2)= G_{AB} s_2^A s_2^B = \sum_{i = 1}^n (s_2^i)^2 d_i +
        h_{\alpha\beta} s_2^{\alpha} s_2^{\beta} \geq 0.
       \eeq
       Using this inequality, (\ref{A.9}),
       $(U^{\Lambda},U^{\Lambda}) < 0$ and $(s,s) =  - 1/(U,U)$
       we get
       \beq{A.12}
        (s,U^{\Lambda})^2 = [ \frac{(U,U^{\Lambda})^2}{(U,U)} -
        (U^{\Lambda},U^{\Lambda})] [(U,U)^{-1} + (s_2,s_2)] >
        \frac{(U,U^{\Lambda})^2}{(U,U)^2} > 0,
      \eeq
       that is equivalent to the  inequality (\ref{3.4}). Thus, (\ref{3.4})
       is proved.

 \begin{center}
 {\bf Acknowledgments}
 \end{center}

  This work was supported in part by the Russian Foundation for
 Basic Research grants Nr. 05-02-17478 and Nr. $07-02-13624-ofi_{ts}$.

\end{document}